\documentstyle[11pt,amssymb,epsf]{article}

\def\ben{\begin{equation}}
\def\een{\end{equation}}

  \let\n=\nu

\let\C=\Chi

\def\nn{\nonumber} \def\bd{\begin{document}} \def\ed{\end{document}}
\def\ds{\documentstyle} \let\fr=\frac \let\bl=\bigl \let\br=\bigr
\let\Br=\Bigr \let\Bl=\Bigl
\let\bm=\bibitem
\let\na=\nabla
\let\pa=\partial \let\ov=\overline
\newcommand{\be}{\begin{equation}}
\newcommand{\ee}{\end{equation}}
\def\ba{\begin{array}}
\def\ea{\end{array}}
\def\ft#1#2{{\textstyle{{\scriptstyle #1}\over {\scriptstyle #2}}}}
\def\fft#1#2{{#1 \over #2}}
\def\del{\partial}
\def\vp{\varphi}
\def\sst#1{{\scriptscriptstyle #1}}
\def\oneone{\rlap 1\mkern4mu{\rm l}}
\def\td{\tilde}
\def\wtd{\widetilde}
\def\ie{\rm i.e.\ }
\def\dalemb#1#2{{\vbox{\hrule height .#2pt
        \hbox{\vrule width.#2pt height#1pt \kern#1pt
                \vrule width.#2pt}
        \hrule height.#2pt}}}
\def\square{\mathord{\dalemb{6.8}{7}\hbox{\hskip1pt}}}
\newcommand{\ho}[1]{$\, ^{#1}$}
\newcommand{\hoch}[1]{$\, ^{#1}$}
\newcommand{\bea}{\begin{eqnarray}}
\newcommand{\eea}{\end{eqnarray}}
\newcommand{\ra}{\rightarrow}
\newcommand{\lra}{\longrightarrow}
\newcommand{\Lra}{\Leftrightarrow}
\newcommand{\bp}{\tilde \beta^\prime}
\newcommand{\tr}{{\rm tr} }
\newcommand{\Tr}{{\rm Tr} }
\def\0{{\sst{(0)}}}
\def\1{{\sst{(1)}}}
\def\2{{\sst{(2)}}}
\def\3{{\sst{(3)}}}
\def\4{{\sst{(4)}}}
\def\5{{\sst{(5)}}}
\def\6{{\sst{(6)}}}
\def\7{{\sst{(7)}}}
\def\8{{\sst{(8)}}}
\def\n{{\sst{(n)}}}
\def\cA{{{\cal A}}}
\def\cB{{{\cal B}}}
\def\cF{{{\cal F}}}
\def\cH{{{\cal H}}}
\def\tV{\widetilde V}
\def\tW{\widetilde W}
\def\tH{\widetilde H}
\def\tE{\widetilde E}
\def\tF{\widetilde F}
\def\tA{\widetilde A}
\def\im{{i}}
\def\tY{{{\wtd Y}}}
\def\ep{{\epsilon}}
\def\vep{{\varepsilon}}
\def\R{\rlap{\rm I}\mkern3mu{\rm R}}
\def\bD{{{\bar D}}}

\def\R{\rlap{\rm I}\mkern3mu{\rm R}}
\def\bD{{{\bar D}}}
\def\R{{{\Bbb R}}}
\def\C{{{\Bbb C}}}
\def\H{{{\Bbb H}}}
\def\CP{{{\Bbb C}{\Bbb P}}}
\def\RP{{{\Bbb R}{\Bbb P}}}
\def\Z{{{\Bbb Z}}}
\def\bA{{{\Bbb A}}}
\def\bB{{{\Bbb B}}}
\def\bC{{{\Bbb C}}}
\def\bD{{{\Bbb D}}}
\def\bE{{{\Bbb E}}}
\def\bZ{{{\Bbb Z}}}
\def\Re{{{\frak{Re}}}}
\def\Im{{{\frak{Im}}}}
\def\cosec{{\,\hbox{cosec}\,}}
\def\Gm{{\Gamma_{\!\! -}}}
\def\Gp{{\Gamma_{\!\! +}}}
\def\stan{{standard }}
\def\nonstan{{supernumerary }}

\newcommand{\tamphys}{\it Center for Theoretical Physics,
Texas A\&M University, College Station, TX 77843}

\newcommand{\upenn}{\it Department of Physics and Astronomy,\\ University
of Pennsylvania, Philadelphia, PA 19104}

\newcommand{\brussels}{\it Physique Th\'eorique et Math\'ematique,
Universit\'e Libre de Bruxelles,\\ Campus Plaine C.P. 231, B-1050
Bruxelles, Belgium}

\newcommand{\auth}{M. Cveti\v c\hoch{*1},  
H. L\"u\hoch{\dagger\ddagger23} and C.N. Pope\hoch{\dagger2}}

\begin{document}
\begin{flushright}
MIFP-04-12\ \ \ UPR-1083-T\ \ \ USTC-ICTS-04-15\\
{\bf hep-th/0406196}\\
June\  2004
\end{flushright}

\vspace{10pt}

\begin{center}

{\large {\bf  Charged Kerr-de Sitter Black Holes in Five Dimensions}}

\vspace{20pt}
\auth

\vspace{20pt}{\hoch{*}\it Department of Physics and Astronomy,\\
University of Pennsylvania, Philadelphia, PA 19104, USA}

\vspace{10pt}{\hoch{\dagger}\it George P. and Cynthia W. Mitchell
Institute for Fundamental Physics,\\ Texas A\& M University,
College Station, TX 77843-4242, USA}

\vspace{10pt} {\hoch{\ddagger}\it Interdisciplinary Center for
Theoretical Study, \\ University of Science \& Technology of China,
Hefei, Anhui 230026, China}

%
%
%

\vspace{40pt}

\underline{ABSTRACT}
\end{center}

    We construct a general class of non-extremal charged Kerr-de
Sitter black holes in five dimensions, in which the two rotation
parameters are set equal.  There are three non-trivial parameters,
namely the mass, charge and angular momentum.  All previously-known
cases, supersymmetric and non-supersymmetric, that have equal angular
momenta are encompassed as special cases.

{\vfill\leftline{}\vfill \vskip 10pt \footnoterule {\footnotesize
\hoch{1} Research supported in part by DOE grant
DE-FG02-95ER40893, NSF grant INT03-24081, and the\\
$\phantom{Xxxx}$ Fay R. and Eugene L.
Langberg Chair.}\vskip 2pt
{\footnotesize
\hoch{2} Research supported in part by DOE grant
DE-FG03-95ER40917.} \vskip 2pt
{\footnotesize \hoch{3} Research supported in part by
grants from the Chinese Academy of Sciences and
the grants from the\\
$\phantom{Xxxx}$ NSF of China with Grant No: 10245001,
90303002.
}

\pagebreak


    Charged black holes with non-zero cosmological constant provide
important gravitational backgrounds for testing the AdS/CFT
correspondence \cite{mal,gkp}. In particular, for charged black holes
in the anti-de Sitter background, the black hole charge plays the role
of the R-charge \cite{cg} in dual field theory. In addition, the
thermodynamic stability as well as analogs of the Hawking-Page phase
transition for such configurations shed light \cite{cg,cg2,cejm} on
the phase structure of the strongly coupled dual field theory.  First
examples of non-extremal charged black holes in five dimensions, as
solutions of a gauged supergravity theory, were obtained in
\cite{bcs}. (For generalizations to other dimensions and their higher
dimensional origin as spinning branes see \cite{10a} .)

    An important generalisation of static charged black holes is to
allow for the rotation.  While the four-dimensional charged rotating
black hole solutions with non-zero cosmological constant, the
Kerr-Newman-de Sitter metrics, were found long ago \cite{carter},
analogs of five-dimensional solutions have not been constructed yet.
General five-dimensional rotating charged black holes with two
non-equal rotation parameters in the zero cosmological constant
background were obtained in \cite{cvetyoum} by employing generating
techniques associated with the underlying non-compact duality
symmetries.  Five-dimensional uncharged rotating black holes with
non-zero cosmological constant, the five-dimensional Kerr-de Sitter
metrics, were obtained a few years ago in \cite{hawhuntay}.  In
addition, certain five-dimensional extremal charged rotating solutions
with non-zero cosmological constant have been found
\cite{klemm1,klemm2,reall}, however, the non-extremal generalisations
have not yet been obtained.  

The purpose of this letter is to present
the solutions for general charged rotating five-dimensional black
holes with a cosmological constant, in the special case where the two
rotation parameters are equal.  To be precise, we consider solutions
of the coupled Einstein-Maxwell system that arises as the bosonic
sector of minimal gauged five-dimensional supergravity, described by
the Lagrangian
\be
{\cal L} = \sqrt{-g}\Big( R -12\lambda - \ft14 F^2 + \fft1{12\sqrt3}\, 
\ep^{\mu\nu\rho\sigma\lambda}\, F_{\mu\nu}\, F_{\rho\sigma}\, 
A_\lambda\Big)\,.\label{boslag}
\ee

     Specialisations of our solutions encompass all the
previously-known cases mentioned above. The general case has three
independent non-trivial parameters, comprising the mass, the charge
and the angular momentum.

   The solutions that we have obtained take the following form:
\bea
ds_5^2 &=& \!\!\!\!-\Big( 1 -\Sigma\, \lambda r^2 -\fft{2M}{r^2} + 
        \fft{Q^2}{r^4}\Big)\, dt^2 + \fft{dr^2}{W} +
 r^2 \, d\Omega_3^2\nn\\
&& - \fft{J^2}{r^4}\, [Q^2 -2(M+Q)\, r^2]\,
(\sin^2\theta\, d\phi + \cos^2\theta\, d\psi)^2\nn\\
&& \!\!\!\!\!\! - 2J\,
\Big(\lambda\, \beta\, r^2  + \fft{2M+Q}{r^2} - \fft{Q^2}{r^4}\Big)\, 
dt\, (\sin^2\theta\, d\phi
+ \cos^2\theta\, d\psi)\,,\label{4paramet}\\
A &=& \fft{\sqrt3  Q}{r^2}\,[ dt -
J\, (\sin^2\theta\, d\phi
+ \cos^2\theta\,  d\psi)]\,.\label{4paravec}
\eea
where
\bea
W &=& 1 -\lambda\, r^2   - \Big[2M + 2\lambda\, J^2\, (M+Q) 
   -2\lambda \, J^2 (2M+Q)\, \beta + 2\lambda^2\, J^4\, (M+Q)\, \beta^2
    \Big]\, \fft1{r^2} \nn\\
&& + \Big[(\lambda\, \beta\, J^2-1)^2\, Q^2 + J^2\, (\lambda\, Q^2 +
    2(M+Q))\Big]\, \fft{1}{r^4}\,,\\
\Sigma &=& 1+ \lambda\, \beta^2\, J^2\,,\label{wsig}
\eea
and 
\be
d\Omega_3^2 = d\theta^2 + \sin^2\theta\, d\phi^2 + \cos^2\theta\, 
   d\psi^2
\ee
is the metric on the unit 3-sphere.  There are three non-trivial parameters,
which may be taken to be $(M,J, Q)$, which are related to the
mass, angular momentum and charge. We shall remark further on the
(trivial) parameter $\beta$ below.

   For some purposes it is convenient to rewrite the metric
(\ref{4paramet}) in terms of left-invariant 1-forms $\sigma_i$ on
$S^3$.  Defining
\bea
\sigma_1 &=& \cos\td \psi\, d\td \theta + \sin\td\psi\, \sin\td\theta
\,d\td\phi\,,\nn\\
\sigma_2 &=& -\sin\td \psi\, d\td \theta + \cos\td\psi\, \sin\td\theta
\,d\td\phi\,,\nn\\
\sigma_3&=& d\td\psi + \cos\td\theta\, d\td\phi\,,
\eea
where 
\be
\psi -\phi = \td \phi\,,\qquad
\psi +\phi = \td \psi\,,\qquad \theta = \ft12 \td\theta\,,
\ee
then we have
\bea
d\theta^2 + \sin^2\theta\,d\phi^2 + \cos^2\theta\, d\psi^2 &=&
\ft14 (\sigma_1^2 + \sigma_2^2 + \sigma_3^2)\,,\nn\\
\sin^2\theta\,d\phi + \cos^2\theta d\psi &=&
\ft12 \sigma_3\,.
\eea
The metric (\ref{4paramet}) can be written as
\be
ds^2 = - \fft{r^2\, W\, dt^2}{4 b^2} + \fft{dr^2}{W} + 
\ft14 r^2\, (\sigma_1^2 + 
\sigma_2^2) + b^2\, (\sigma_3 + f\, dt)^2\,,\label{4paramet2}
\ee
where
\bea
b^2 &=& \ft14 r^2\, \Big( 1 -\fft{J^2\, Q^2}{r^6} + \fft{2J^2\, 
(M+Q)}{r^4}\Big)
\,,\nn\\
f&=& -\fft{J}{2b^2}\, \Big( \lambda\, \beta\, r^2 + \fft{2M+Q}{r^2} 
    -\fft{Q^2}{r^4}\Big)\equiv \fft{U}{2b^2}\,.
\eea
The potential (\ref{4paravec}) becomes
\be
A = \fft{\sqrt3 Q}{r^2}\, (dt - \ft12 J\, \sigma_3)\,.
\ee
Note that the function $W$, given in (\ref{wsig}), can be written as
\be
W = \fft{4b^2}{r^2}\, (f^2 \, b^2 -g_{00})=
\fft{1}{r^2}\, (U^2 - 4 b^2\, g_{00})\,,\label{simpleW}
\ee
where $g_{00}$ is the coefficient of $dt^2$ in (\ref{4paramet}).

   As we remarked earlier, the parameter $\beta$ is trivial.  Here, we 
demonstrate this in the case when $Q=0$.\footnote{We are grateful to 
Gary Gibbons and
Malcolm Perry for discussion that led to this conclusion.  More
recently, it has been shown in \cite{madros} that $\beta$ is also
trivial when $Q\ne0$.}  To demonstrate its triviality when
$Q=0$, the key observation is that one can perform the coordinate 
transformation
\be
\td\psi \longrightarrow \td\psi + c\, t\label{psitrans}
\ee
in the metric (\ref{4paramet2}), which therefore has the effect of 
shifting the function $f$ by the additive constant $c$:
\be
f\longrightarrow f + c\,.\label{ftrans}
\ee
This allows us to eliminate $\beta$ as an independent parameter, when
$Q=0$.  To see this, we introduce the constant $k$ by
\be
k\equiv \fft{ \sqrt{1+ 4\, \lambda\, \beta\, {J'}^2} -1}{2\lambda\, \beta\,
   {J'}^2}\,,
\ee
in terms of which we define new mass and angular momentum parameters
via
\be
M =M'/k^2,\qquad J =k\, J'\,.
\ee
It is easy to see that the metric functions $W$, $b$ and $f$ are now
expressed as
\bea
W &=& 1-\lambda\, r^2 - \fft{2M'\, (1+\lambda\, {J'}^2)}{r^2} 
    + \fft{2M'\, {J'}^2}{r^4}\,,\\
b^2 &=& \ft14 r^2\, \Big(1 + \fft{2M'\, {J'}^2}{r^4}\Big)\,,\\
f &=& \fft{1-\sqrt{1+4\lambda\, \beta\, {J'}^2}}{J'} - 
      \fft{4 M'\, J'}{r^4 + 2 M'\, {J'}^2}\,.\label{fnew}
\eea
Using the constant shift transformation (\ref{ftrans}) induced by the
coordinate transformation (\ref{psitrans}), we see that the first term
in (\ref{fnew}) can be removed, and hence $\beta$ no longer appears in
the metric.  This proves that $\beta$ is a trivial parameter, in this
situation when $Q=0$.  See \cite{madros} for a proof that $\beta$ is
trivial also when $Q\ne0$.  It is nonetheless useful to retain the
redundant parameter $\beta$, since this provides a convenient way to
consider various limits.

\medskip\bigskip
\noindent{{\bf Reductions to Previously-known Solutions:}}
\medskip

\begin{itemize}

\item[1.]  In the case where the charge $Q$ vanishes, the solutions reduce to
those of Hawking, Hunter and Taylor-Robinson \cite{hawhuntay}, in the
special case of equal angular momenta in the two orthogonal transverse
2-planes.  To be precise, if we send $r^2 \rightarrow (r^2
+J^2)/\Sigma$, $M\rightarrow M/\Sigma$, $J\rightarrow a/\Sigma$ and choose
$\beta=\Sigma$, then the metric (\ref{4paramet}) reduces to that given
in \cite{hawhuntay}, with their rotation parameters $a$ and $b$ set equal.

\item[2.]  If instead we take the cosmological constant $\lambda$ to
vanish, and set
\be
\beta=c+s\,,\quad Q= 2\mu\, s\, c\,, \quad M=\mu\, (c^2+s^2)\,,\quad
J=(c-s)\, \ell\,,\quad
r^2\rightarrow  r^2 + \ell^2 + 2\mu\, s^2
\ee
where $c\equiv \cosh\delta$ and $s\equiv \sinh\delta$,
then the solutions (\ref{4paramet}) and (\ref{4paravec}) reduce to a
special case of those found  in  \cite{cvetyoum}, in which the two
angular momenta are set equal, $\ell_1=\ell_2=\ell$, and the three
charges of the more general Einstein-Maxwell-Dilaton considered there
are set equal.  As the parameter $\delta$ is increased from 0 to
$\infty$, with an accompanying inverse scaling of $\mu$ so that the
charge $Q$ and mass remain finite, the solutions interpolate between
zero-charge Kerr black holes and supersymmetric extremal black holes.
       
\item[3] One can obtain BPS metrics with $\lambda\ne0$ by taking
$Q=\pm M$.  These encompass the previously-known BPS solutions.
Specifically, we find:
\bea
\hbox{Klemm-Sabra \cite{klemm1}}:&& Q= -M\,,\quad \beta=0\,,\\
\hbox{Gutowski-Reall \cite{reall}}:&& Q=M\,,\quad 
J= \ft12\sqrt{-\lambda}\, M\,,
\quad \beta = -\fft{2}{\lambda\, M}\,.\label{gutreall}
\eea
Note that the Gutowski-Reall case, for which $\Sigma\equiv 1+\lambda\,
\beta^2\, J^2 =0$, requires that the cosmological constant $\lambda$
be negative.  The constant $M$ is denoted by $r_0^2$ in \cite{klemm1}.
In the Klemm-Sabra case, the constants $M$ and $J$ are denoted by $q$
and $-a/q$ in \cite{klemm1}.

\end{itemize}

   The Gutowski-Reall solution and the Klemm-Sabra solution are both
supersymmetric, within the minimal ${\cal N}=2$ supergravity whose
bosonic sector is described by (\ref{boslag}).  We can obtain a more
general class of supersymmetric solutions with $Q=M$, generalising
\cite{reall}, by imposing only one additional condition, namely that
$\beta^2\, J^2 = -1/\lambda$.  (Although trivial as a parameter in
general, here $\beta$ effectively parameterises a family of
inequivalent BPS limits.)

\medskip\bigskip
\noindent{\bf Global Analysis:}
\medskip

    Horizons occur where the $r^2\, W/(4 b^2)$ prefactor of the $dt^2$
term in (\ref{4paramet2}) vanishes.  Closed time-like curves (CTCs)
occur if $b^2<0$, and so for a regular black hole with no naked CTCs,
we require that $W(r)$ vanish at some positive value of $r$ for which
$b^2$ is still positive as $r$ reduces from infinity (or from the
cosmological horizon if $\lambda>0$).  It follows from (\ref{simpleW})
that naked CTCs can arise if, for example, $|Q|> M$ and $\Sigma >0$.
If $|Q|$ is sufficiently small in comparison to $M$, these regularity
conditions will be satisfied for a range of values of the parameter
$\beta$.  Here we present a comprehensive study of the conditions
on the four parameters for a regular black hole in which there are no
naked singularities or CTCs.

      The properties of the general solutions are largely determined
by the functions $W$ and $b^2$.  The solutions are invariant under
$r\leftrightarrow -r$, with a curvature singularity at the fixed point
$r=0$.  Thus without loss of generality we need consider only the
region $r\ge 0$. Here we shall consider in detail only the solutions
with a negative cosmological constant.

For solutions to be free of naked singularities, there should exist an
event horizon at $r=r_+>0$, which is the largest root of $W$.  For the
solutions to be free of naked CTCs, we require that $b_+^2 \equiv
b(r_+)^2 >0$, and that $b^2$ remain positive for all $r>r_+$.  The
entropy and the temperature of the solution are then given by
\be
T=\fft{r W'}{8\pi\, b}\Big|_{r=r_+}\,,\qquad
S=\pi^2\, r^2\, b\Big|_{r=r_+}\,.\label{tempent}
\ee
For our solutions, $r_+^2$ is indeed the largest root of $W$ when we
have $W'(r_+) \ge 0$.  Thus, for a well-defined regular black hole, we
have the following conditions:
\be
r_+>0\,,\qquad W(r_+)=0\,,\qquad b_+^2 >0\,,\qquad
W'(r_+)\ge 0\,.\label{gencon0}
\ee
For our solutions, the minima of $r^4\, b^2$ occur at $r<r_+$, thus
the condition $b^2_+>0$ also ensures that $b^2>0$ for all $r>r_+$.  To
simplify the analysis, it is convenient to define dimensionless
quantities $(m,q,j,\alpha)$ by
\be
M=m\,r_+^2\,,\qquad Q=q\, r_+^2\,,\qquad
J=j\, r_+\,,\qquad \lambda = -\fft{\alpha^2}{r_+^2}\,.
\ee

   The global structure of the uncharged solutions, which reduce to those
in \cite{hawhuntay} with $a=b$, is well established. 
In order to study the global structure of the general solutions
with $Q\ne0$, let us assume that there exists an event horizon at $r_+>0$,
and then we have
\bea
W(r_+)=&&-(2m + 2q -q^2) (1 + \alpha^2\beta\,j^2)^2 + 2 
\alpha^2 q\,j^2\,\beta
+2 (1 + \alpha^2) (m+q)\,j^2+\nn\\&& + 1 +
\alpha^2 + 2 q - \alpha^2 j^2 q^2 =0\,.
\label{qbeta}
\eea
This is a quadratic equation for $\beta$.  The existence of a solution
for a real value of $\beta$ requires that the discriminant, which is given 
by
\be
\Delta = \fft{16\alpha^4 b_+^2 (4 b_+^2 - r_+^2)^2}{
r_+^6 (2m + 2q - q^2)^2}\, \Bigl(2 (1 + \alpha^2)(m+q) - \alpha^2 q^2\Bigr)
\,,
\ee
be non-negative.  Here $b_+$ is given by
\be
b_+^2 =\ft14 r_+^2\, (1 + j^2 (2m + 2 q - q^2))\,.
\ee
Thus we have
\be
2 (1 + \alpha^2)(m+q) - \alpha^2 q^2\ge 0\,,\qquad
1 + j^2 (2m + 2 q - q^2)>0\,.\label{genqcon1}
\ee
For $r_+$ to be the event horizon, we need further to require that
$r_+^2$ be the largest root of $W$.  Assuming this to be the case, the
temperature and entropy are given by
\bea
S&=&\ft12\pi^2\, r_+^3\,\sqrt{1  + j^2 (2m + 2q -q^2)}\,,\nn\\
T&=&\ft{\pi\,r_+^2}{4(2m+2q -q^2)\,S}\,\Big(-4j^2 (m+q)^2 +
2 (1 +2\alpha^2)(m+q)\nn\\
&&\qquad\qquad - (2 + 3 \alpha^2)q^2 - 2(1 + \alpha^2 j^2\beta) q^3
\Big)\,,
\eea
where $\beta$ is given by (\ref{qbeta}).  The requirement that $r_+$
be the largest root of $W$ provides a further constraint that the
right hand side of the $T$ above is non-negative, in addition to
constraints given in (\ref{genqcon1}).

        Since $W$ is a cubic polynomial in the
variable $r^2$, there might arise three positive roots for $r^2$.  In
this case, $W'$ for the smallest root would also be positive,
seemingly satisfying the above conditions.  To examine this, we denote
the other two roots by $\td r_\pm^2$, and we find that
\be
r_+^2 - \td r_\pm^2
=\fft{(1 + 3\alpha^2)\,r_+^2}{2\alpha^2}\pm
\fft{r_+\sqrt{\pi (1+3\alpha^2)^2\, r_+^2 - 16 \alpha^2\,
T\, S}}{2\alpha^2\,
\sqrt{\pi}}\,.
\ee
If $TS > \pi (1 + 3\alpha^2) r_+^2/(16\alpha^2)$, then the function
$W$ has only one real root for $r^2$, namely $r_+^2$.  If on the other
hand, $0\le TS \le \pi (1 + 3\alpha^2) r_+^2/(16\alpha^2)$, there can
be two additional roots, but $r_+^2$ is the largest one.  In
particular, when $T=0$, $r_+$ and $\td r_-$ coincide, implying that
$W$ has a second-order zero, as one would expect.

       Note that  the temperature and entropy calculation above 
becomes singular when $m$ and $q$ are related by $m=\ft12 q^2 - q$.
In this case, we have
\be
b_+^2=\ft14 r_+^2 >0\,.
\ee
In fact it is easy to see that $b^2$ is positive-definite for $r\ge
r_+$, implying the absence of naked CTCs.  The existence of a horizon
requires that
\be
\beta=-\fft{1 + \alpha^2 + 2q + j^2 q^2}{2 \alpha^2 j^2 q}
\,.
\ee
The temperature and the entropy are given by
\be
S=\ft12 \pi^2\,r_+^3\,,\qquad
T=\fft1{8\pi\,r_+} \Big(
-j^4 q^4 - 2(3-\alpha^2) j^2 q^2 -\alpha^4 +6\alpha^2 +3\Big)\,.
\ee
Note that in this case, the entropy is independent of any dimensionless
parameters. The requirement $W'(r_+)\ge 0$ implies that
\be
 -3 - 2 \sqrt3 + \alpha^2 \le j^2 q^2 \le -3 + 2 \sqrt3 + \alpha^2\,.
\ee
The case of extremality, $T=0$, occurs at the equalities of the above.
Note that as a function of $j^2\, q^2$, the temperature lies in the
range $0\le T \le 3/(2 \pi\, r_+)$.

   In the above analysis, we started with the conditions
(\ref{gencon0}), with $\beta$ being arbitrary.  For special solutions
such as those in \cite{hawhuntay} ($\beta=1$) and \cite{reall}
($\beta=1/(\alpha j)$), it would become cumbersome to extract the
conclusions from our final results.  In these cases, it is more
convenient to discuss directly the condition (\ref{gencon0}), with the
specific $\beta$ values substituted in.

       The existence of Killing spinors in these solutions does not in
general preclude the possibility of having naked CTCs.  The $Q=-M$
solutions, which include the Klemma-Sabra specialisation when
$\beta=0$, have naked CTCs for all values of $\beta$.  This can be
easily seen from the fact that when $Q=-M$, we have
\be
W=\fft{(r^2 - (1 - \lambda\,\beta\, J^2)Q)^2}{r^4}-4\lambda\, b^2\,.
\ee
Thus, for negative cosmological constant $\lambda$, the horizon, $W=0$, 
occurs for  $b^2<0$, only.

        For the other branch of the supersymmetric solution, where
$M=Q$ and $\Sigma=0$, we have $(-g_{00})= (1 -Q/r^2)^2$, which is
non-negative.  It follows from (\ref{simpleW}) that the horizon at
$W=0$ occurs where $b^2<0$, unless $U=0$ at $r^2=Q$.  This condition
is uniquely satisfied by the Gutowski-Reall solution (\ref{gutreall}).
The conditions for extremality ($T=0$) and supersymmetry of our
solutions do not necessarily coincide.  The supersymmetric solutions
we discussed above have two parameters $Q$ and $J$, and these black
holes in general have non-vanishing temperature.  This result
contradicts the assumption about the stability of configurations that
is normally associated with supersymmetry.  However, as we saw above,
the supersymmetric black holes with non-vanishing temperature all have
naked CTCs.  The one that does not have naked CTCs, found in
(\ref{gutreall}), indeed has zero temperature.

To conclude, we have obtained a general class of five-dimensional de
Sitter charged rotating black hole solutions in which the two angular
momenta are set equal.  The solutions depend on three non-trivial
parameters, naemly the mass, the angular momentum and the
charge.  All previously-known cases,
extremal and non-extremal, supersymmetric or non-supersymmetric, that
have equal momenta are encompassed as special cases.  We analysed the
conditions under which one
obtains black holes with no naked singularities or closed time-like
curves.  We also found that the supersymmetry condition does not in
general imply zero temperature.  However, the supersymmetric solutions
with non-vanishing temperature all have naked closed time-like curves.
The only one \cite{reall} with no naked closed time-like curves indeed
has zero temperature.

    The general class of charged AdS$_5$ black holes presented in this
paper provides a fertile ground to further study their thermodynamics,
stability and phase transitions as well as having other implications
for the AdS$_5$/CFT$_4$ correspondence. In particular, it would be
interesting to study further the connection between thermodynamics and
the closed time-like curves from the point of view of $D=4$
super-conformal field theory.

Further generalizations of such rotating configurations to two
non-equal rotation parameters as well as to three non-equal charges
of five-dimensional gauged supergravity with $U(1)$ gauged isometry,
are other important directions \cite{clp}.  In addition, it would be
interesting to generalise these solutions to charged Kerr-de Sitter
solutions in higher dimensions, thus extending the general spinning
charged black holes with zero cosmological constant \cite{cy2} and the
general Kerr-de Sitter solutions, obtained recently in \cite{glpp}, to
charged black holes in de Sitter backgrounds.

\section*{Acknowledgment}

We are grateful to Zhi-Wei Chong, Gary Gibbons and Malcolm Perry for useful
discussions.  We are grateful to the Relativity Group in DAMTP,
Cambridge (M.C. and C.N.P.) and CERN theory Division (M.C.)  and ICTS
at USTC, Heifei (H.L.)  for the hospitality.

\section*{Note Added}

   In an earlier version of this paper (the one published
in Phys. Lett. {\bf B598}, 273 (2004)), it was wrongly stated that the
parameter $\beta$ was non-trivial when $Q\ne0$.  This has recently 
been corrected by S.F. Ross and O. Madden \cite{madros}.

\end{document}